\documentclass[prc,aps,groupedaddress,preprintnumbers,onecolumn]{revtex4}

\usepackage{graphicx} 
\usepackage{amssymb}
\usepackage{epstopdf}
\usepackage{hyperref}
\usepackage{epsfig}
\usepackage{multirow}
\usepackage{verbatim}
\usepackage{url}

\newcommand{\antinue}{\ensuremath{\overline{\nu}_{e}}}
\begin{document}

\title{Multiple Detectors for a Short-Baseline Neutrino Oscillation Search Near Reactors}

\author{K.M.~Heeger }
\affiliation
        {\it Department of Physics,
University of Wisconsin, Madison, WI 53706, USA}
\affiliation
        {\it Physics Department, Yale University,
New Haven, CT 06511, USA}
\author{B.R.~Littlejohn}
\email{littlebh@uc.edu}
\affiliation
        {\it Physics Department,
University of Cincinnati, Cincinnati, OH 45221, USA}
\author{H.P.~Mumm}
\affiliation
        {\it National Institute of Standards and Technology, Gaithersburg, MD 20899, USA}
\date{\today}


\begin{abstract}
{Reactor antineutrino experiments have the ability to search for neutrino oscillations independent of reactor flux predictions using a relative measurement of the neutrino flux and spectrum across a range of baselines. The range of accessible oscillation parameters are determined by the baselines of the detector arrangement. We examine the sensitivity of short-baseline experiments with more than one detector and discuss the optimization of a second, far detector. The extended reach in baselines of a 2-detector experiment will improve sensitivity to short-baseline neutrino oscillations while also increasing the ability to distinguish between 3+1 mixing and other non-standard models.}
\end{abstract}

\maketitle

\section{Introduction}
\label{sec:introduction}

Anomalous results from a variety of neutrino experiments and astrophysical observations have been interpreted as an indication of new physics in the neutrino sector\cite{MB, LSND, Gallium2, Astro}. In particular, a re-evaluation of reactor flux predictions has led to an increase in the predicted reactor $\antinue$ flux by about 3.5\% \cite{HuberAnomaly, Anomaly2} while the spectral shape and its uncertainties remain largely unchanged. When compared with experimental data at baselines between 10-100\,m these calculations suggest a $\sim$6\% difference between the measured and predicted reactor antineutrino flux~\cite{Anomaly1}.  This ``{reactor anomaly}'' can be interpreted as either a sign of new physics or due to uncertainties in the reactor $\antinue$ flux predictions. It has been suggested that such a deficit could be due to the oscillation of an additional sterile neutrino state with a mass splitting of the order of $\sim 1{\rm eV}^2$ and a corresponding oscillation length of $\mathcal{O}$(3\,m) \cite{Anomaly1}.  It has also been noted that the addition of multiple sterile neutrinos can help mitigate the apparent tension between the appearance oscillation results from MiniBooNE and LSND and null disappearance results from other experiments~\cite{Giunti, Kopp}.  Additional searches for oscillations at short-baselines are thus well motivtaed to test the ``reactor anomaly'' and search for the possible existence of light sterile neutrinos. 

A direct approach for investigating the apparent discrepancy in the reactor neutrino fluxes is to search for the characteristic $L/E$ neutrino oscillation pattern by measuring the reactor $\antinue$ spectrum at distances comparable to the expected sterile neutrino oscillation length of $\mathcal{O}$(3\,m) \cite{AnomalyWhite}.  Such a measurement has the ability to search for neutrino oscillations independent of the reactor flux predictions.  Recent work \cite{VSBL} explored the experimental parameters necessary for a short-baseline (SBL) reactor experiment and demonstrated that a majority of the suggested oscillation parameter space can be tested at high confidence level~\cite{VSBL}. Past work suggests the construction of a radially extended detector with good energy and position resolution at a high-power, compact-core research reactor facility. This allows the precision measurement of the spectral shape over a range of baselines. High event statistics and  good position (L) and energy (E) resolution extend the sensitivity reach to low values of $\theta_{new}$. A measurement over a range of baselines increases the tested parameter space in $\Delta m^2$. This approach is now utilized by a number of proposed short-baseline reactor experiments worldwide~\cite{Jianglai, Serebrov, NIST, STEREO}.

In this work we suggest increasing the range of available baselines and optimized sensitivity through the use of multiple detectors.  Although briefly mentioned in Refs.~\cite{VSBL} and~\cite{Jianglai}, a detailed exploration of a multi-detector SBL oscillation experimental concept has not been performed.  The purpose of this paper is to demonstrate the increased physics reach of a short-baseline experiment with 2 detectors and to explore the design parameters of the additional `far' detector.  Section~\ref{sec:benefits} introduces the physics case for a multi-detector SBL reactor experiment.  Section~\ref{sec:sig} briefly overviews the input experimental assumptions and analysis method, and then describes the benefits in sensitivity afforded by multi-detector arrangements in both the 3+1 and 3+2 sterile neutrino pictures.  Section~\ref{sec:optimization} explores the optimization of the far detector design in the context of proposed reactor sites and near detectors.

\section{Extending the $L/E$ Range with Multiple Detectors}
\label{sec:benefits}

A demonstration of neutrino oscillation independent of reactor flux predictions relies on the relative measurement of the spectral shape and rate as a function of distance. A radially extended single detector over a suitably long baseline fulfills this function. However, practical considerations such as the available floor space, constraints associated with the reactor confinement building, and access pathways in any cases limit the allowable length of a segmented detector at reactor sites. This work shows that an optimized experimental setup with two or more detectors provides significant improvement in the range of $L/E$ and can achieve sensitivity similar to a single radially extended detector over the baselines of interest. 

Proposed short-baseline experiments at research reactors aim to search for oscillations with a single detector at baselines of $<$10~m ($L/E\approx2.5$). The maximum measurable value of $L/E$ for a two detector experiment depends on the power of the reactor, the size of the detector, and possible detector locations. For some of the US research reactors considered in recent studies \cite{VSBL} placing a second detector at a distance of 15-20~m ($L/E\approx7$) appears feasible, more than doubling the range of measurable $L/E$ values. Doubling the $L/E$ coverage includes the first oscillation maximum for small-$\Delta m^2$ oscillations and allows for the observation of multiple periods for larger $\Delta m^2$. As a result, the overall sensitivity of the experiment is improved and a broader range in $\Delta m^2$ is covered. 

Recent reactor experiments to measure the mixing angle $\theta_{13}$ have used a relative analysis between ``identical'' near and far detectors to aid in the cancellation of detector systematics.  The purpose of a second, far detector in a short-baseline experiment is similar in the sense that it allows for a relative measurement of the flux and spectrum at longer baselines but we emphasize that in this case no special cancellation of relative systematic uncertainties is assumed. A short-baseline oscillation experiment utilizes the relative measurement between sub-segments within a single detector to search for oscillations independent of reactor flux predictions.  The addition of a second, far detector esentially increases the number of ``segments'' and the baseline range, rather than introducing previously unavailable possibilities for cancellation of detector systematics. We also note that the backgrounds to the near and far detectors in a short-baseline experiment will likely be very different.  We cannot make the assumption of identical signal to background in these detectors and will have to treat them as an independent measurement at different baselines. For the purpose of this study we assume no additional cancellation of systematics between detectors beyond that applied between detector sub-segments.  For further discussion of systematics cancellation see Ref. ~\cite{Serebrov, VSBL}.

\section{Experimental Parameters and Sensitivity}
\label{sec:sig}


Table~\ref{tab:params} overviews the key experimental design parameters used for the studies and figures shown in this paper along with relevant references.  Reactor and background parameters are identical to those in Ref.~\cite{VSBL} and we refer therein for detailed explanations. As an extension of this earlier work the detector parameters now describe two detectors, a near and far detector.

\begin{table}[!htb]
  \noindent\makebox[\textwidth]{%
    \begin{tabular}{l | l | l | l | l} \hline
      \multicolumn{2}{c|}{Parameter} & Value & Comment & Reference \\ \hline
      \multirow{4}{*}{Reactor} & Power & 20 MW & NIST-like & \cite{NISTschedule}\\
      & Shape & cylindrical & NIST-like & \cite{NISTschedule} \\
      & Size & 0.5~m radius; 1~m height & NIST-like & \cite{NISTschedule} \\
      & Fuel & HEU & Research reactor fuel type & \cite{NISTschedule, HFIRschedule, ATRschedule}\\ \hline
      \multirow{6}{*}{Detector} & Dimensions, Near & 1.2$\times$0.65$\times$2.1~m & 2.1 meters available baseline & - \\
      & Dimensions, Far & 2.4$\times$3.25$\times$2.1~m & 2.1 meters available baseline & - \\
      & Efficiency & 30\% & In range of SBL exps. (10-50\%) & \cite{Lasserre, Bugey, Rovno, Bowden} \\  
      & Proton density & 6.39$\times$10$^{28}\frac{p}{m^3}$ & In range of LS Exps & \cite{TDR} \\
      & Position resolution & 15~cm & Daya Bay-like & \cite{Bryce}  \\
      & Energy resolution & 10\%/$\sqrt{E}$ & Daya Bay-like & \cite{Bryce} \\ \hline
      \multirow{4}{*}{Background}& \multirow{2}{*}{S:B ratio} & \multirow{2}{*}{1} & In range of SBL exps. (1-25) & \cite{Bugey, Bowden, ILL}\\
      & & & In range of SBL R\&D (1) & \cite{Lasserre}\\
      & \multirow{2}{*}{Background shape} & \multirow{2}{*}{1/E$^2$ + Flat} & Low-Energy Accidentals (1/E$^2$) & \multirow{2}{*}{-} \\
      & & & Neutron Bkg (Flat Approximation) & \\ \hline
      \multirow{4}{*}{Other} & Run Time (nominal) & 1-3 years live-time & -  & - \\ 
      & Closest Distance, Near & 3.75~m & Available at NIST & -  \\
      & Closest Distance, Far & 15.75~m & Available at NIST & -  \\
      & Prompt Energy Threshold & 2.0~MeV & -  & -  \\ \hline
    \end{tabular}}
  \caption{Nominal experimental parameters used for all sensitivity calculations presented in this paper.  The optimization studies in Section~\ref{sec:optimization} utilize these baseline values while varying specific parameters as indicated.  Unless noted otherwise, parameters apply to both near and far detectors.}
  \label{tab:params}
\end{table}

\label{subsec:chisquare}

The sensitivity of a liquid scintillator reactor experiment to neutrino oscillations is evaluated by comparing the detected inverse beta decay prompt events $T_{ij}$ in energy bin $i$ and position bin $j$ to the expected events $M_{ij}$ in the absence of neutrino oscillations and in the presence of a background $B_{ij}$.  For the purposes of these calculations, $T_{ij}$ is taken as $B_{ij}$ plus an oscillated version of $M_{ij}$.  A $\chi^2$ is used to test the hypothesis of no-oscillation and for oscillation parameter estimation in the case of either one or two additional sterile neutrino states, identically to that presented in Ref~\cite{VSBL}:

\begin{eqnarray}
\label{eq:chi2}
\displaystyle
\chi^2 = \sum_{i,j}\frac{\left[M_{ij} - (\alpha + \alpha_e^i + \alpha_r^j)T_{ij} - (1+\alpha_b)B_{ij} \right]^2 }{T_{ij}+B_{ij}+(\sigma_{b2b}B_{ij})^2}  
+\displaystyle \frac{\alpha^2}{\sigma^2} + \displaystyle\sum_{j}\left(\frac{\alpha_r^j}{\sigma_r}\right)^2 + \displaystyle\sum_{i}\left(\frac{\alpha_e^i}{\sigma_e^i}\right)^2 + \frac{\alpha_b^2}{\sigma_b^2}.
 \end{eqnarray}

The $\chi^2$ sum is minimized with respect to the relevant oscillation parameters and to the nuisance parameters \{$\alpha$, $\alpha_r^j$, $\alpha_e^i$, $\alpha_b$\} describing the systematic uncertainties of the measurement, as described in~\cite{PDG}.  These nuisance parameters represent the overall normalization, relative position normalization, uncorrelated energy spectrum, and background systematics. Associated bounding uncertainties of these systematics are \{$\sigma$, $\sigma_r$, $\sigma_b$\} = \{100\%, 0.5\%, 10\%\}.  The uncorrelated energy spectrum uncertainties $\sigma_e^i$ follow the description given in~\cite{HuberAnomaly}.  An additional uncertainty $\sigma_{b2b}$ of 0.5\% is added to the $\chi^2$ to account for uncertainties in the position and energy distribution of backgrounds, which are not currently well-understood, and are likely to be uncorrelated between energy and position bins.  A more pedagogical description of the $\chi^2$ and its components is given in~\cite{VSBL}.  Again we note that no special cancellation of detector systematics between near and far detectors has been assumed in these studies unless explicitly stated.

\subsection{Sensitivity in the 3+1 framework}

The 3+1 neutrino model with one additional sterile neutrino state and a mass splitting of~$\sim$1~eV$^2$ mass is frequently used in the literature to benchmark the sensitivity of new experiments to short-baseline oscillations \cite{AnomalyWhite}.  In keeping with this tradition, we will first present the sensitivity to 3+1 neutrino oscillations for one and two detectors. We note, however, that the underlying  physics is not yet known. A short-baseline experiment will search for oscillatory effects as a signature of new physics. Should a deviation from the expected $1/r^2$ behavior of the reactor $\overline{\nu}_{e}$ be found, short-baseline oscillations can be interpreted in a variety of models including 3+1 neutrino mixing.  In the 3+1 framework, oscillations occur between the three known neutrino flavors and a single additional sterile neutrino state of widely differing mass.  The short-baseline $\overline{\nu}_e$ survival probability associated with this oscillation is described by

\begin{eqnarray}
\label{eq:Eq31}
P_{ee} = 1-4|U_{e4}|^2(1-|U_{e4}|^2)\sin^2\frac{\Delta m^2 L}{4E} = 1 - \sin^22\theta_{ee}\sin^2\frac{\Delta m^2 L}{4E},
 \end{eqnarray}

with the oscillation amplitude $\sin^22\theta_{ee}$ = $\sin^22\theta_{14}$ = 4$|U_{e4}|^2(1-|U_{e4}|^2)$.
 
\begin{figure}[htb!pb]
\centering
\includegraphics[trim=0.1cm 1cm 0.1cm 1cm, clip=true, width=0.95\textwidth]{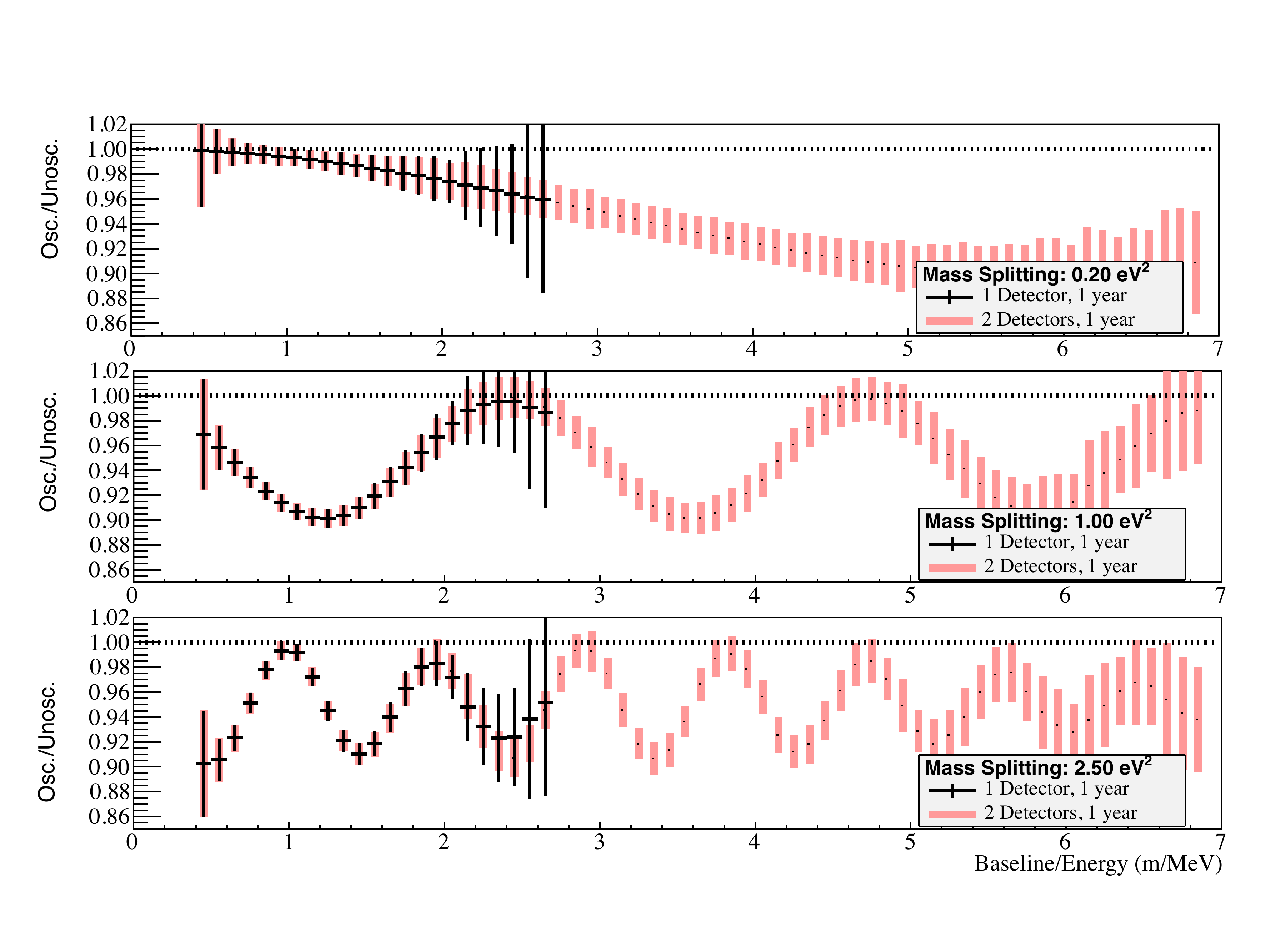}
\includegraphics[trim=0.1cm 0.1cm 0.1cm 0.1cm, clip=true, width=0.6\textwidth]{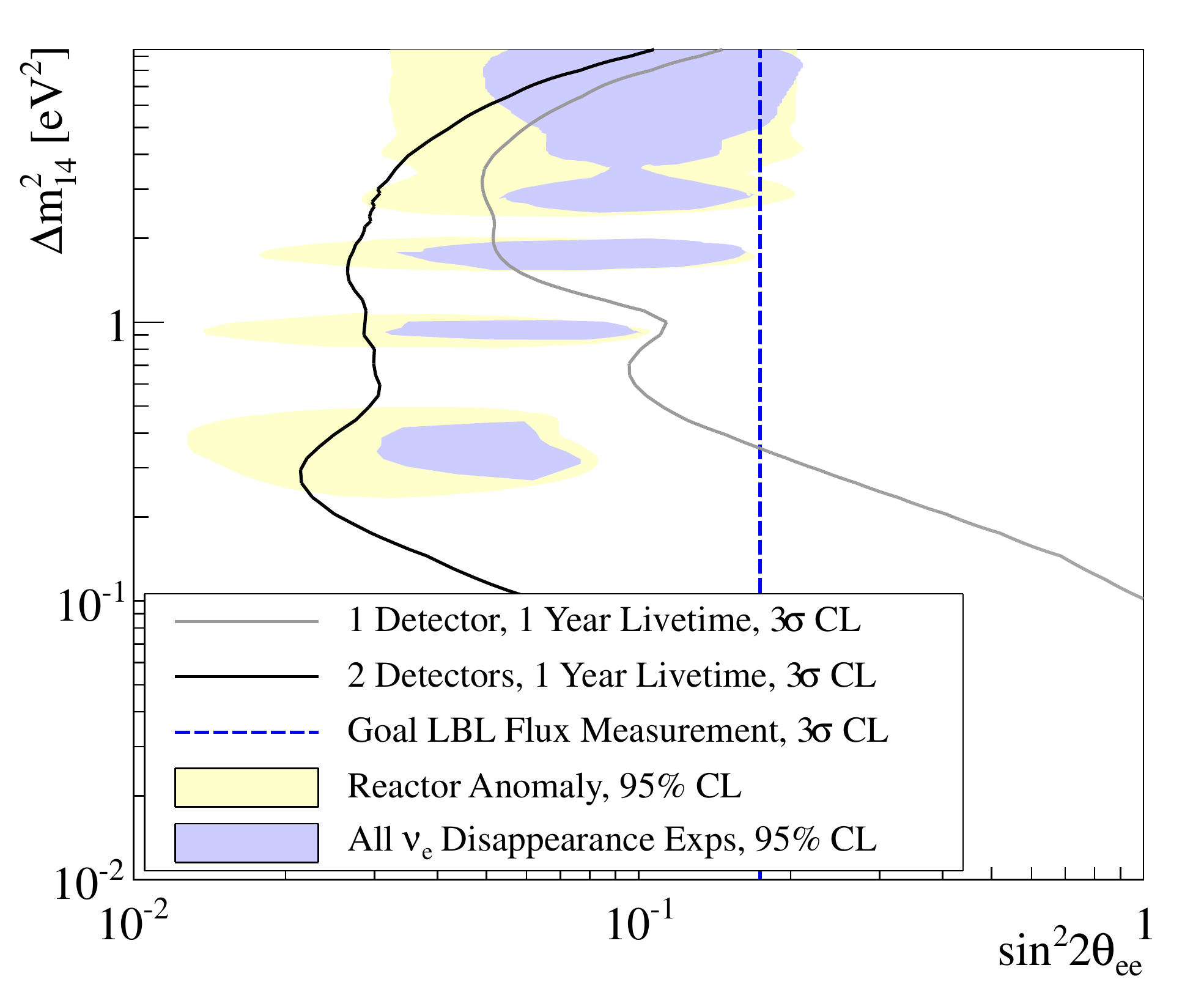}
\caption[]{Top: Oscillated $L/E$ distributions for 3+1 neutrino mixing assuming an oscillation amplitude of 10\% and 1 year of data with one or two detectors. The default parameters described in Table~\ref{tab:params} are used.  In the two-detector case, the far detector active target mass is 10$\times$ that of the near detector.  Bottom: Total sensitivity to 3+1 oscillations for these 1- and 2-detector arrangements.  The far detector active volume is 10$\times$ the size of the near detector.  Statistics increase from roughly 160,000 detected events with one detector to 270,000 detected events with two detectors.  The 2-detector configuration is capable of ruling out the majority of the 95\% CL region suggested by $\nu_e$ disappearance experiments. The vertical dashed line indicates the expected sensitivity of longer baseline $\overline{\nu}_e$ disappearance measurements of the reactor $\theta_{13}$ experiments. }
\label{fig:Osc31}
\end{figure}

Figure~\ref{fig:Osc31} shows oscillated $L/E$ distributions assuming the existence of one sterile neutrino state for 1 year of data with one or two detectors  at three values of $\Delta m^2$.  The measured $L/E$ distributions include smearing from the finite experimental position reconstruction and the energy resolutions shown in Table~\ref{tab:params}.  As a second detector is added, the $L/E$ coverage increases from around 2-3~m/MeV to greater than 6~m/MeV.  

This increase in $L/E$ coverage translates to an increase in the observable range of $\Delta m^2$ values.  In particular, at low $\Delta m^2$ values, sensitivity is greatly improved as the long oscillation wavelength are resolved within the experiment's wider $L/E$ coverage.  One can also see an increase in sensitivity for  $\Delta m^2$ values covered by a one-detector arrangement.  At medium values of $\Delta m^2$, additional $L/E$ coverage allows the detection of multiple neutrino oscillation periods, rather than a single or partial period.  The additional reach in $L/E$ increases the ability to distinguish any observed oscillation from the null hypothesis of no-oscillation.  At high $\Delta m^2$, the total number of visible oscillation periods is also increased, although the finite experimental resolution tends to damp out the oscillation effect. The resultant increase in the sensitive range of $\Delta m^2$ and $\theta_{14}$ going from one to two detectors is illustrated in Figure~\ref{fig:Osc31}.

\subsection{Distinguishing 3+2 Neutrino Mixing and Other Non-Standard Physics}
Interpretation of data taken in a short-baseline experiment in terms of 3+1 mixing is the simplest model-dependent description of an oscillatory signal.  In reality, the observed experimental signature may be due to a more complex physics scenario. This section explores the sensitivity of a 2-detector experiment to 3+2 neutrino mixing and its ability to  distinguish  3+1 mixing from 3+2, or other non-standard models.  We will utilize the oscillation formalism and parametrization described in Ref.~\cite{Kopp}. In the 3+2 framework, oscillations occur between the three known neutrino flavors and two additional much heavier sterile neutrino states.  Oscillation also occurs between the two sterile neutrino states.  This gives a total $\overline{\nu}_e$ survival probability of:

\begin{eqnarray}
\label{eq:Eq32}
P_{ee} = 1-\left(1-\sum_{i=4,5}|U_{ei}|^2\right)\sum_{i=4,5}|U_{ei}|^2 \sin^2\frac{\Delta m_{i1}^2 L}{4E} - 4|U_{e4}|^2|U_{e5}|^2\sin^2\frac{\Delta m_{54}^2 L}{4E}.
 \end{eqnarray}
We use the $\chi^2$ described in Section~\ref{subsec:chisquare} to determine the experiment's ability to exclude oscillations resulting from two sterile states for a range of 3+2 oscillation parameters. A larger range of accessible $L/E$ values obviously increases the experiment's ability exclude the no-oscillation hypothesis and to discriminate against 3+1 models. 

\begin{figure}[htb!pb]
\centering
\includegraphics[trim=3.5cm 2cm 13.5cm 2.2cm, clip=true, width=0.46\textwidth]{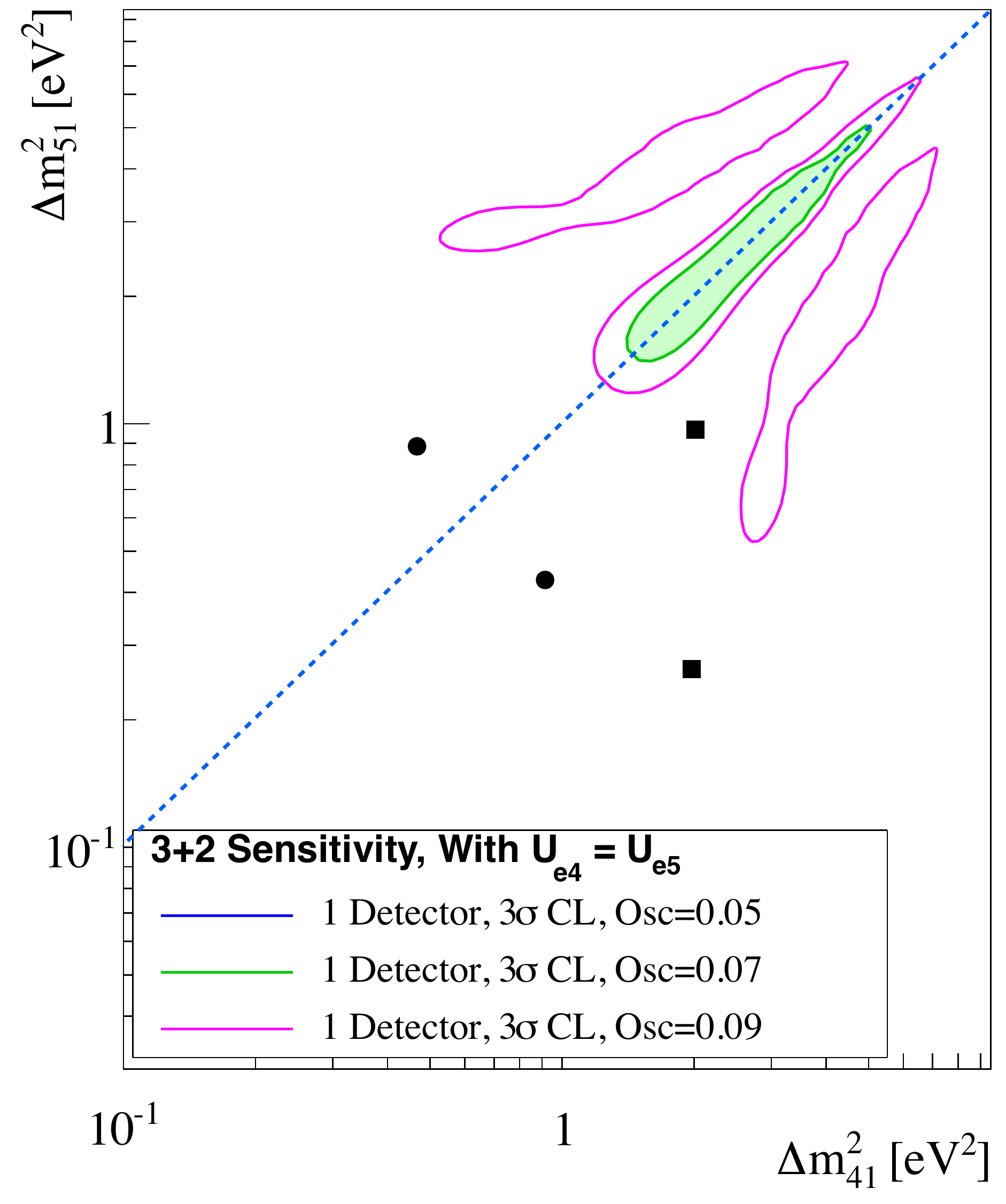}
\includegraphics[width=0.46\textwidth]{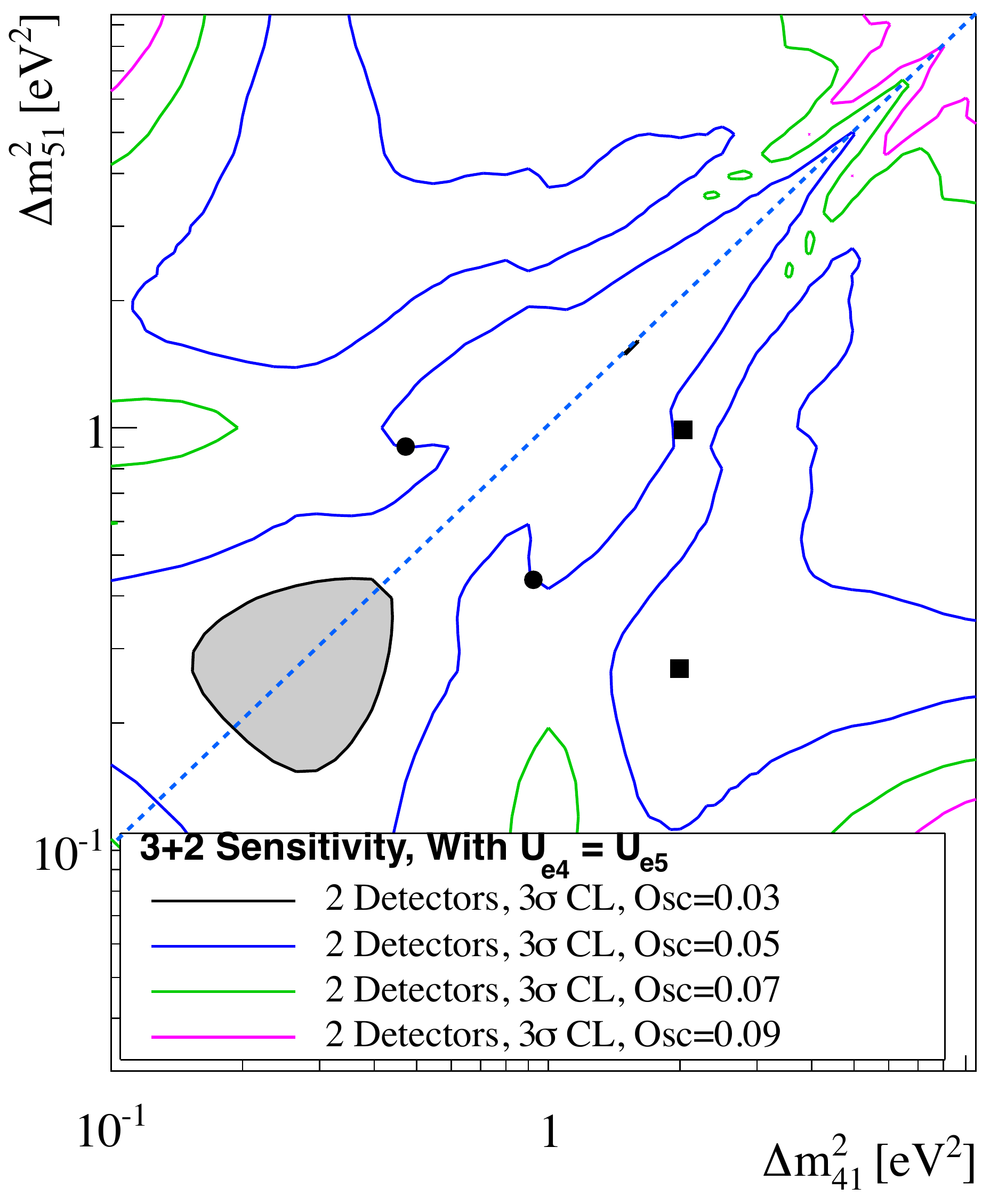}
\caption[]{3+2 oscillation exclusion regions in $\Delta m^2_{41}$ vs $\Delta m^2_{51}$ for 1 year of data from one (left) or two (right) detectors with default parameters as described in Table~\ref{tab:params}.  We assume equal oscillation amplitudes to each of the two sterile states ($|U_{e4}|^2$ = $|U_{e5}|^2$) while varying the total oscillation amplitude.  This assumption roughly matches the best-fit 3+2 oscillations in Ref.~\cite{Kopp}.  The shaded regions in the figures illustrate the excluded 3+2 parameter space for a specific choice of oscillation amplitude.  The off-diagonal exclusion regions show the constraints on the 3+2 model under the given assumptions. The square markers serve as reference points for Figures~\ref{fig:LE32} and~\ref{fig:LE32_Comp}, while the circles indicate the best-fit 3+2 mass squared splittings from~\cite{Kopp}.
}
\label{fig:Osc32}
\end{figure}


Figure~\ref{fig:Osc32} shows exclusion contours in the 3+2 model for varying maximum oscillation amplitudes given the simplifying condition of identical mixing with the two sterile states ($|U_{e4}|^2 = |U_{e5}|^2$) and for a one-detector (left) and two-detector (right) arrangements.  For reference, the diagonal lines represent $\Delta m^2_{51}$ = $\Delta m^2_{41}$ and are analogous to the case of 3+1 mixing. The regions off the diagonal are the regions of non-degenerate 3+2 oscillations. The complete confidence level contours have four degrees of freedom to properly represent the four relevant 3+2 oscillation parameters, thus a multi-dimensions  global fit is needed to generically constrain the allowable parameter space.  

\begin{figure}[htb!pb]
\centering
\includegraphics[width=0.85\textwidth]{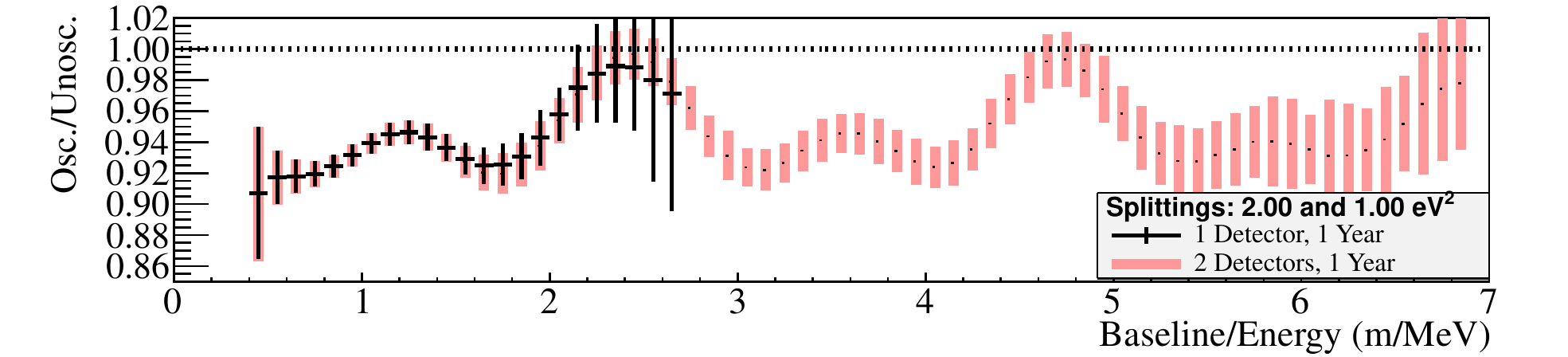}
\includegraphics[width=0.85\textwidth]{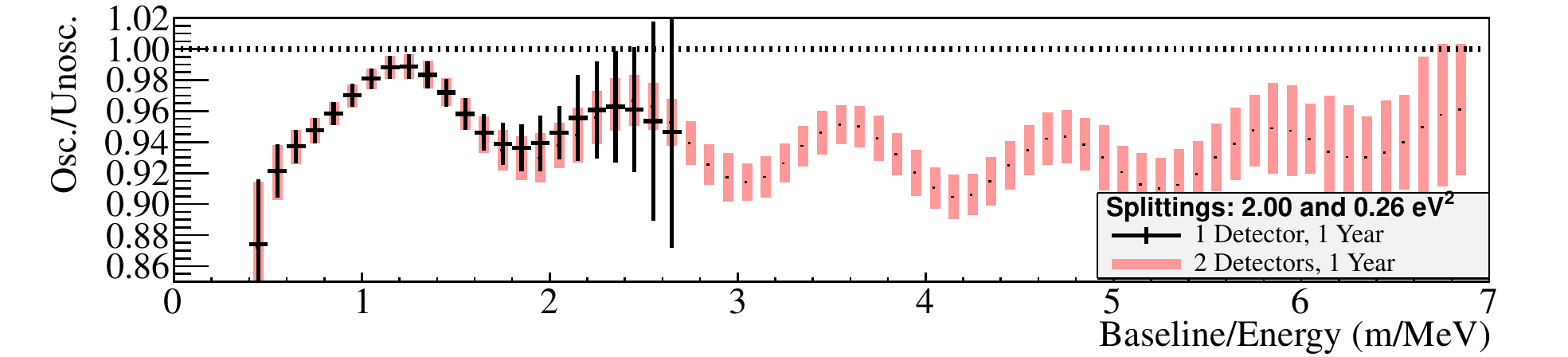}
\caption[]{Oscillated $L/E$ distributions in 3+2 models from 1 year of data with one or two detectors, default parameters as described in Table~\ref{tab:params}, and assuming the existence of two sterile neutrino states with mass-squared splittings of 2~eV$^2$ and 1~eV$^2$ (top) and 2~eV$^2$ and 0.26~eV$^2$ (bottom) respectively.  These mass splittings are chosen to illustrate the interference between the two oscillation effects.  The overall amplitude is reduced compared to 3+1 mixing resulting in a decreased sensitivity to discover oscillations. The sensitivity  to exclude 3+2 oscillations at these mass splitting combinations is shown in Figure~\ref{fig:Osc32}.}
\label{fig:LE32}
\end{figure}

For models with multiple sterile neutrino states interference between oscillation patterns reduces the overall oscillation amplitude for particular $L/E$ values and reduces the experiment's sensitivity to discover short-baseline oscillation. Figure~\ref{fig:LE32} illustrates this for two combinations of $
\Delta m^2$.  When one mass-squared splitting is twice the other (top), the first oscillation maximum, which has the highest statistics is maximally reduced in amplitude, allowing for a better no-oscillation fit to the full oscillation pattern, and less sensitivity to 3+2 mixing.  Alternatively, if one mass-squared splitting is only a fraction of the other (bottom), oscillations occur on different scales, and interference has little negative effect on the sensitivity.  Figure~\ref{fig:Osc32} also shows that off-diagonal sensitivity is best when the two sterile mass-squared splittings are similar to the two best ranges of sensitivity along the diagonal: one splitting around 0.26 eV$^2$ and the other from 2-4 eV$^2$ for the particular experimental arrangement investigated in this paper.  As in the 3+1 case, the range of accessible mass splittings is also greatly increased for two detectors over one-detector experimental arrangements.



\begin{figure}[htb!pb]
\centering
\includegraphics[width=0.85\textwidth]{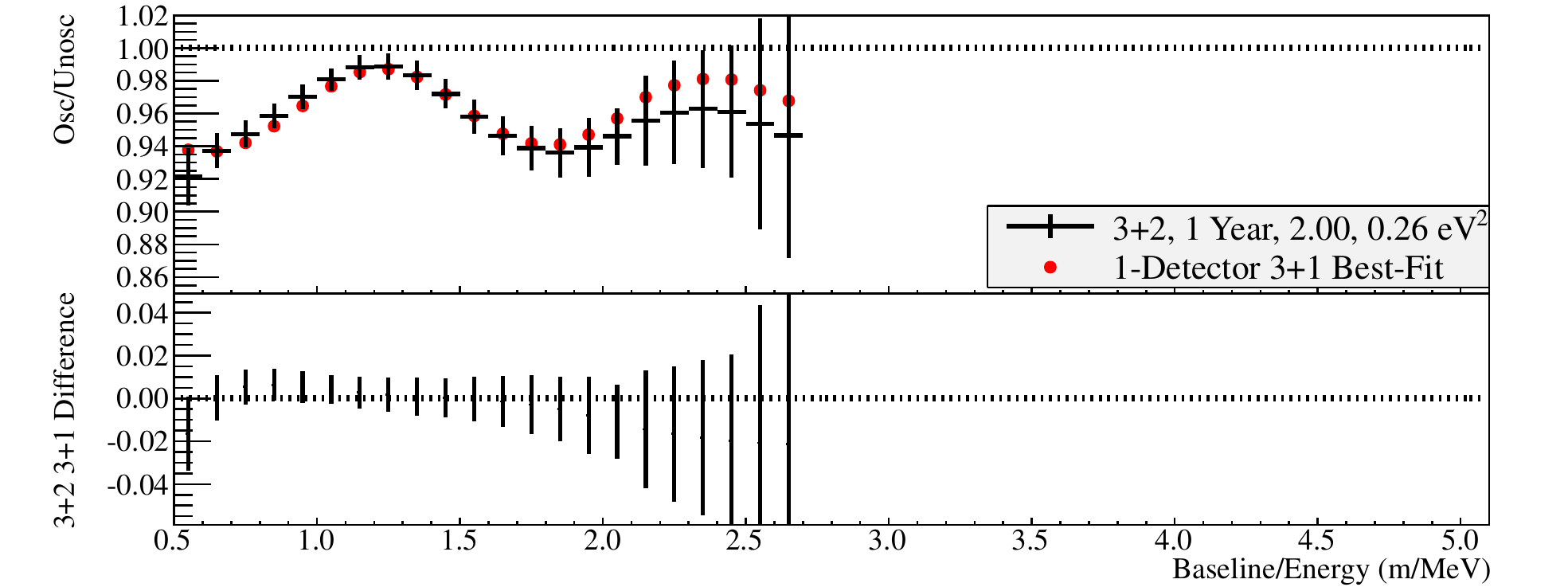}
\includegraphics[width=0.85\textwidth]{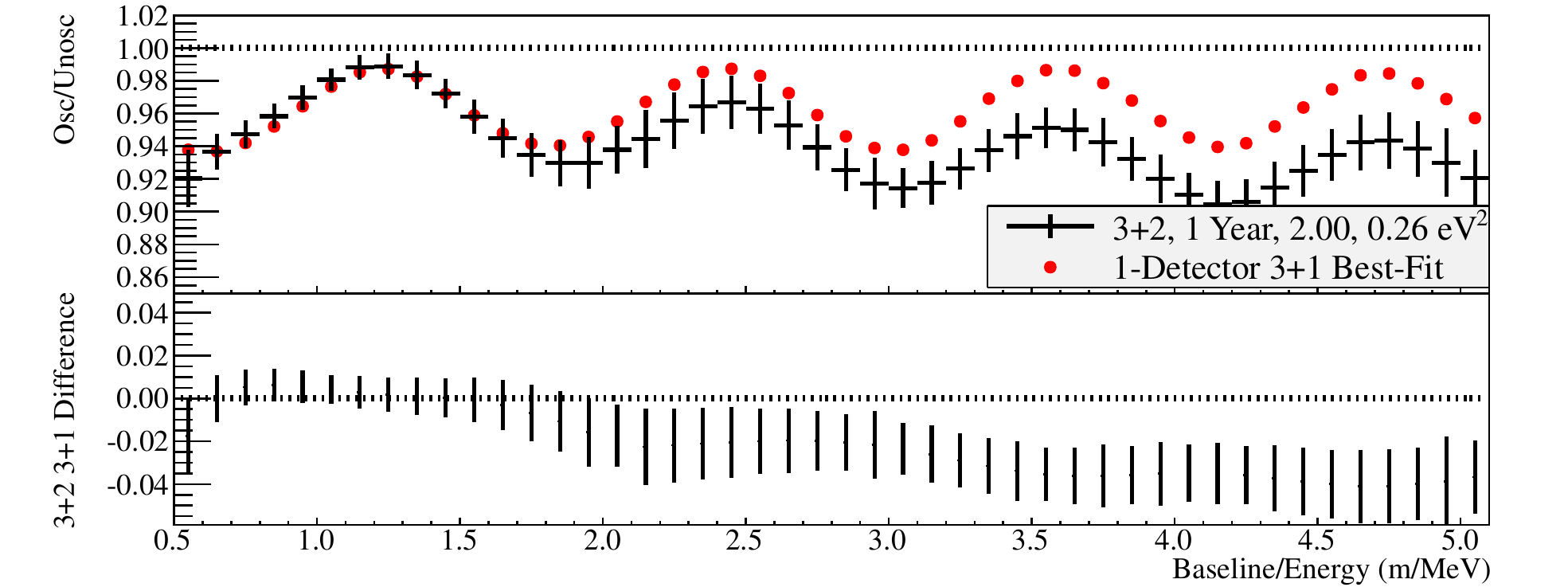}
\caption[]{Simulated 3+2 oscillated $L/E$ distributions from 1 year of data of one (top) or two (bottom) detectors with default parameters from Table~\ref{tab:params}, assuming a 10\% oscillation amplitude and $|U_{e4}|^2$ = $|U_{e5}|^2$.  Also plotted on both panels is the  3+1 oscillation best-fit to the one-detector data.  The best-fit includes a free-floating overall normalization (0.99 times the true normalization for the best-fit) and no constraints on the oscillation parameters (which float to $\sin^22\theta_{14}$ = 0.055 and $\Delta m^2_{41} = 2.0$~eV$^2$.  The lower halves of the panels show fractional difference between the 3+2 simulated data and the 3+1 best-fit.  Adding the second detector to the experimental configuration reveals the inconsistency of the data with the 3+1 fit. 
}
\label{fig:LE32_Comp}
\end{figure}

\begin{figure}[htb!pb]
\centering
\includegraphics[trim=0.5cm 1.5cm 15cm 2.5cm, clip=true, width=0.5\textwidth]{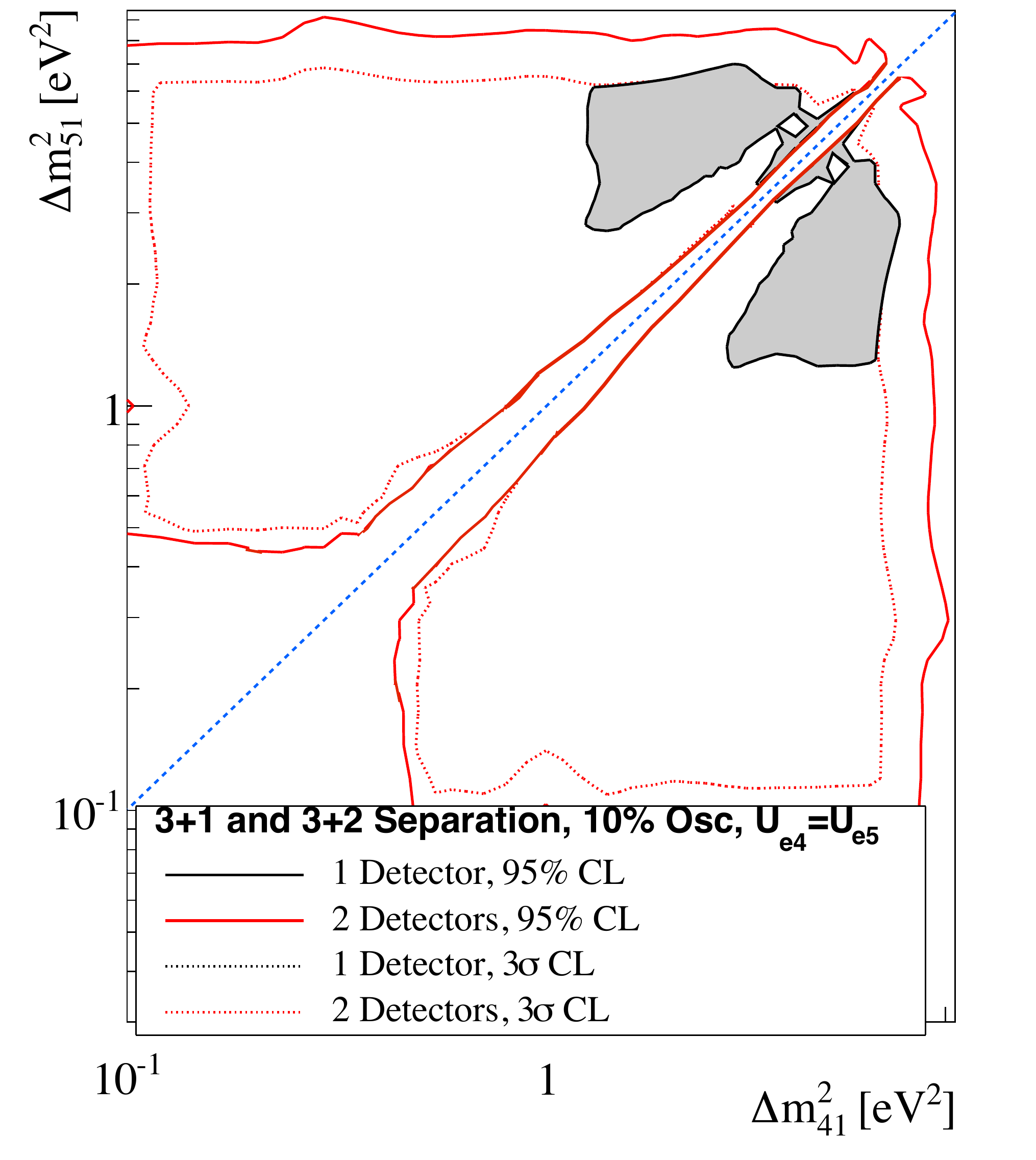}
\caption[]{Exclusion regions describing the ability to distinguish between 3+2 and 3+1 sterile neutrino mixing at 95\% or 3$\sigma$ confidence level for the default one- and two-detector experimental arrangement.  For illustrative and simplifying purposes, a total oscillation amplitude of 10\% is used, and we assume $U_{e4}=U_{e5}$.  If one (two) light sterile neutrino exists, the plot defines the areas where the 3+2 (3+1) oscillation hypothesis can be ruled out along with the no-oscillation hypothesis.  As indicated by the shading in the figure, the regions inside the contour are excluded.  Given a constant total oscillation amplitude, these excluded regions will naturally contract as $U_{e4}$ shrinks relative to $U_{e5}$.  Note that a 1-detector experimental configuration has no ability to distinguish between 3+2 and 3+1 models at 3$\sigma$ CL.}
\label{fig:31V32}
\end{figure}

The ability to discriminate between 3+1 and 3+2 mixing models can be tested by exercising the $\chi^2$ defined in section~\ref{subsec:chisquare} as was done in the 3+2 mixing case, while implementing the 3+1 oscillation for $M_{ij}$, rather than using no oscillation.  In this process, the 3+1 $\Delta m^2$ and $|U_{e4}|$ parameters are allowed to vary independently of the 3+2 mixing parameters.  The result of this minimization for the case of equal 3+2 mixing amplitudes and a maximum oscillation amplitude of 10\% is shown in Figure~\ref{fig:31V32}.  One can see that the addition of a second detector greatly increases the mass-splitting range over which a SBL experiment is able to distinguish between the two oscillation scenarios.  As long as the mass splittings are within the range of those that can be discriminated from no oscillation for the 3+1 case, and as long as the splittings are not too close to one another, one can distinguish between 3+1 and 3+2 oscillation cases at $95\%$ CL.  This sensitivity is naturally reduced if one assumes widely differing values of $|U_{e4}|$ and $|U_{e5}|$. As one mixing parameter gets larger while holding the total oscillation amplitude constant, it becomes harder to distinguish 3+2 mixing from a 3+1 oscillation picture consisting of only the more dominant mass splitting.
It should also be noted that all mass-splitting terms in the 3+2 SBL oscillation equation are squared, meaning that SBL reactor oscillation experiments have no ability to distinguish between 3+2 and 1+3+1 mixing.

\section{Far Detector Optimization in a 2-Detector Experiment}
\label{sec:optimization}

Having established the purpose of a second detector in extending the effective $L/E$ range we now explore the parameters that will optimize its usefulness.
The reactor antineutrino flux falls off as $1/r^2$ as a function of distance from the reactor core. To obtain sufficient event statistics at farther distances the size of the far detector needs to be larger than the near detector. The impact of the far detector on the oscillation sensitivity of the experiment depends on the mass splitting $\Delta m^2$. At high values of $\Delta m^2$ the oscillation can be measured in the near detector alone and the relative impact of the far detector is less. In the range of high $\Delta m^2$ the near-detector statistics and systematics limit the experiment's sensitivity. At low $\Delta m^2$ the far detector measures a suppression in the neutrino flux as shown in Figure \ref{fig:Osc31} and provides a normalization to the overall oscillation effect. In this case the ratio between the near and far detector measurements provides the sensitivity of the experiment.  The experimental parameters of the near detector have been explored before ~\cite{VSBL}.  We now optimize the characteristics of the far detector to maximize the sensitivity of the experiment over a broad range of $\Delta m^2$. 

\subsection{Far Detector Size}

\begin{figure}[htb!pb]
\centering
\includegraphics[trim=0.1cm 0.1cm 0.1cm 0.1cm, clip=true, width=0.49\textwidth]{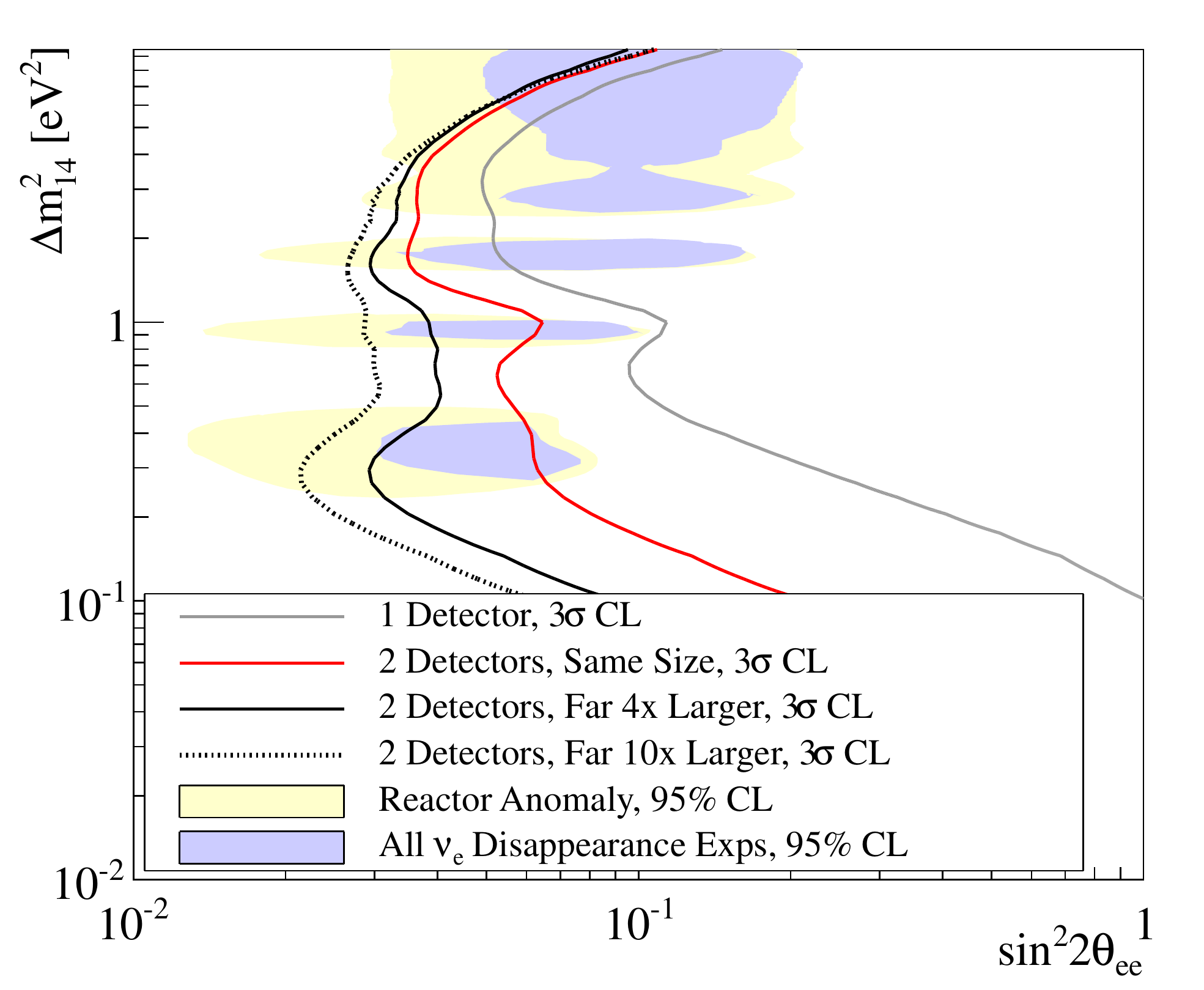}
\includegraphics[trim=0.1cm 0.1cm 0.1cm 0.1cm, clip=true, width=0.49\textwidth]{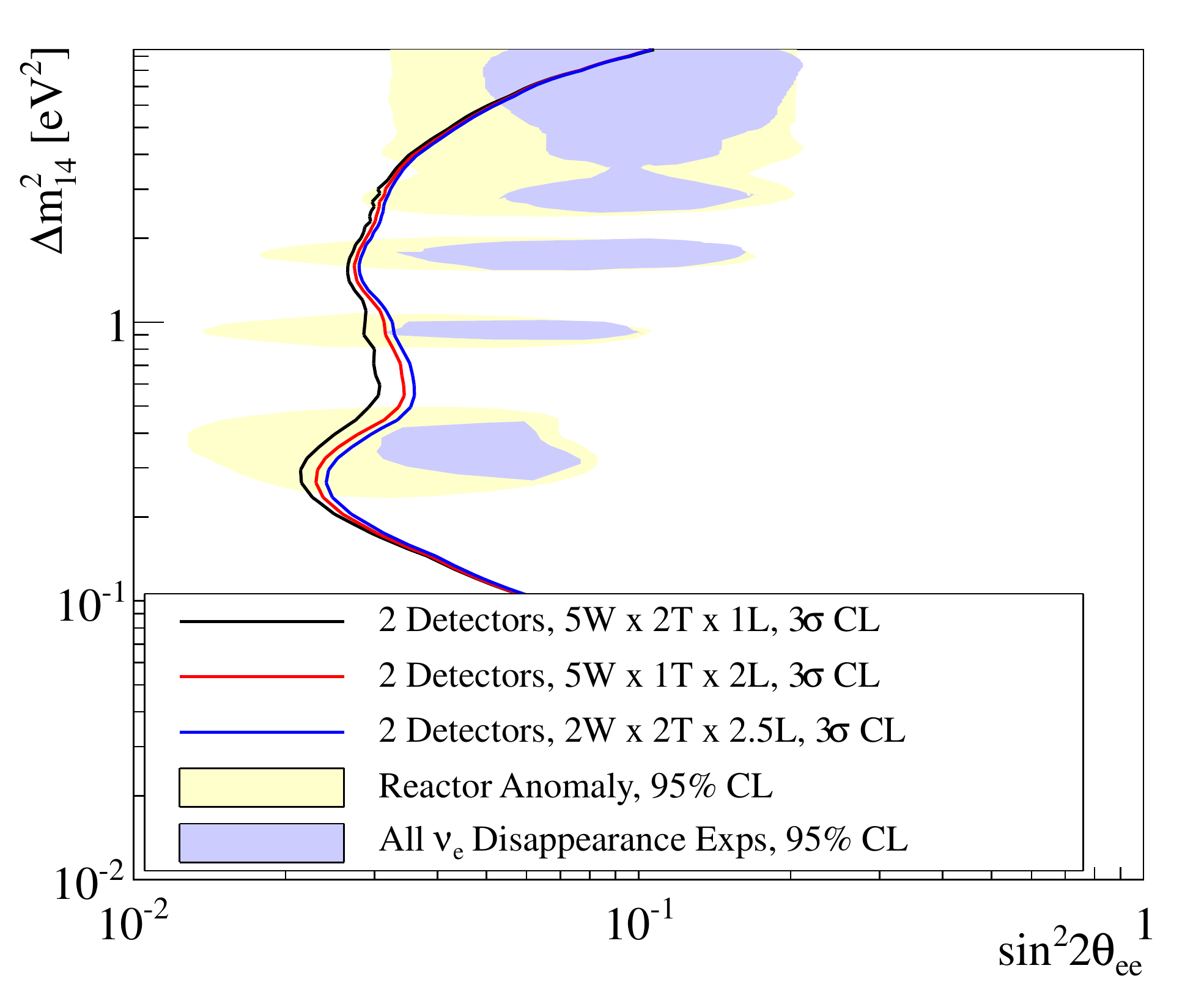}
\caption[]{Sensitivity of the default two-detector experimental arrangement to 3+1 oscillations assuming various far detector sizes (left) and shapes (right).  Sizes of the far detector are given in length of baseline (L), width (W), and height (H), relative to the size of the near detector length, width and height.  Sensitivity is clearly benefited by utilizing a far detector larger in size than the near detector that has minimal spread in baseline.  Sensitivities remain unchanged when detector height and width dimensions are interchanged.}
\label{fig:FarSizeShape}
\end{figure}

As the size of a far detector target is increased, the event statistics at larger $L/E$, as well as resultant experimental sensitivity, is naturally increased.  Figure~\ref{fig:FarSizeShape} shows sensitivity to 3+1 oscillations for various far detector sizes.  Significant gains in sensitivity are visible up to a far detector target ten times the size of the near detector, although the gain in sensitivity is most drastic for the target increased by a factor of 4.  While a sizable far detector is important, scaling of the active detector size by $1/r^2$ is unnecessary. 

\subsection{Far Detector Shape}
\label{subsec:FarShape}

The principal function of the far detector is to provide a normalization of the measured reactor flux at large baselines for low $\Delta m^2$ values.  Maximizing the size of the far detector and the measured event rate at a particular baseline will have the highest impact on providing this normalization. This suggests a  far detector with high cross-sectional area in a small range of radii so that the uncertainty in the baseline is small; an onion-shell like detector would be ideal.  Practically, the shape of the far detector target can be varied either by elongating the detector such that it subtends a broader range of baseline values, or by altering the ratio between the height and width of the detector.  The right panel of Figure~\ref{fig:FarSizeShape} demonstrates changes in sensitivity as these shape variations are applied to the default two-detector arrangement while keeping the shortest distance between reactor and detector target constant.  One can see a reduction in experimental sensitivity as the detector is elongated in baseline while keeping the volume constant.  For a two-detector arrangement with the chosen default parameters, the loss in statistics at a particular baseline outweigh any gain from the additional $L/E$ spreads afforded by an elongated far detector.

\subsection{Reactor to Detector Distance}

\begin{figure}[htb!pb]
\centering
\includegraphics[trim=0.1cm 0.1cm 0.1cm 0.1cm, clip=true, width=0.53\textwidth]{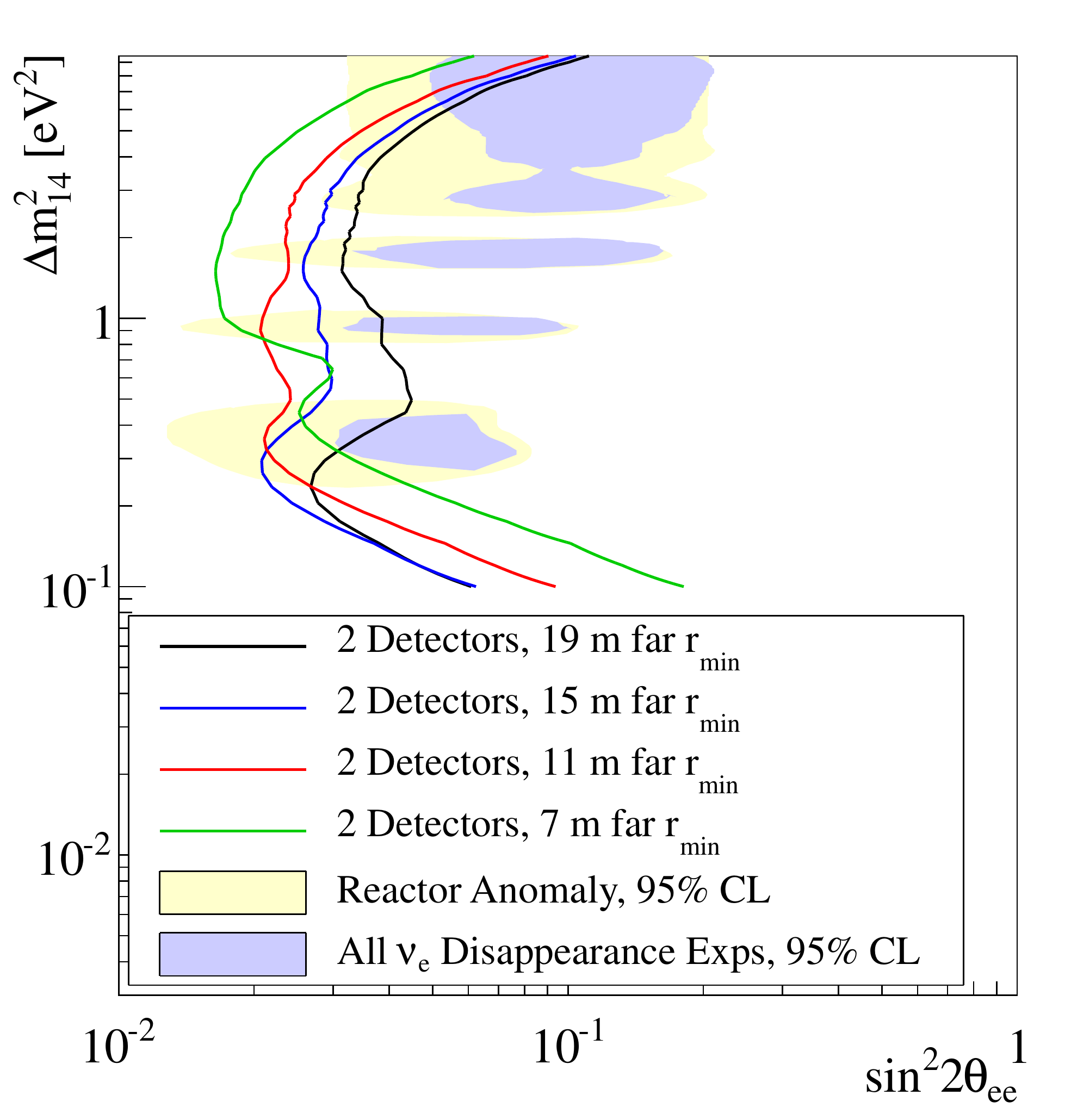}
\includegraphics[trim=2.0cm 0.1cm 14.5cm 4.0cm, clip=true, width=0.44\textwidth]{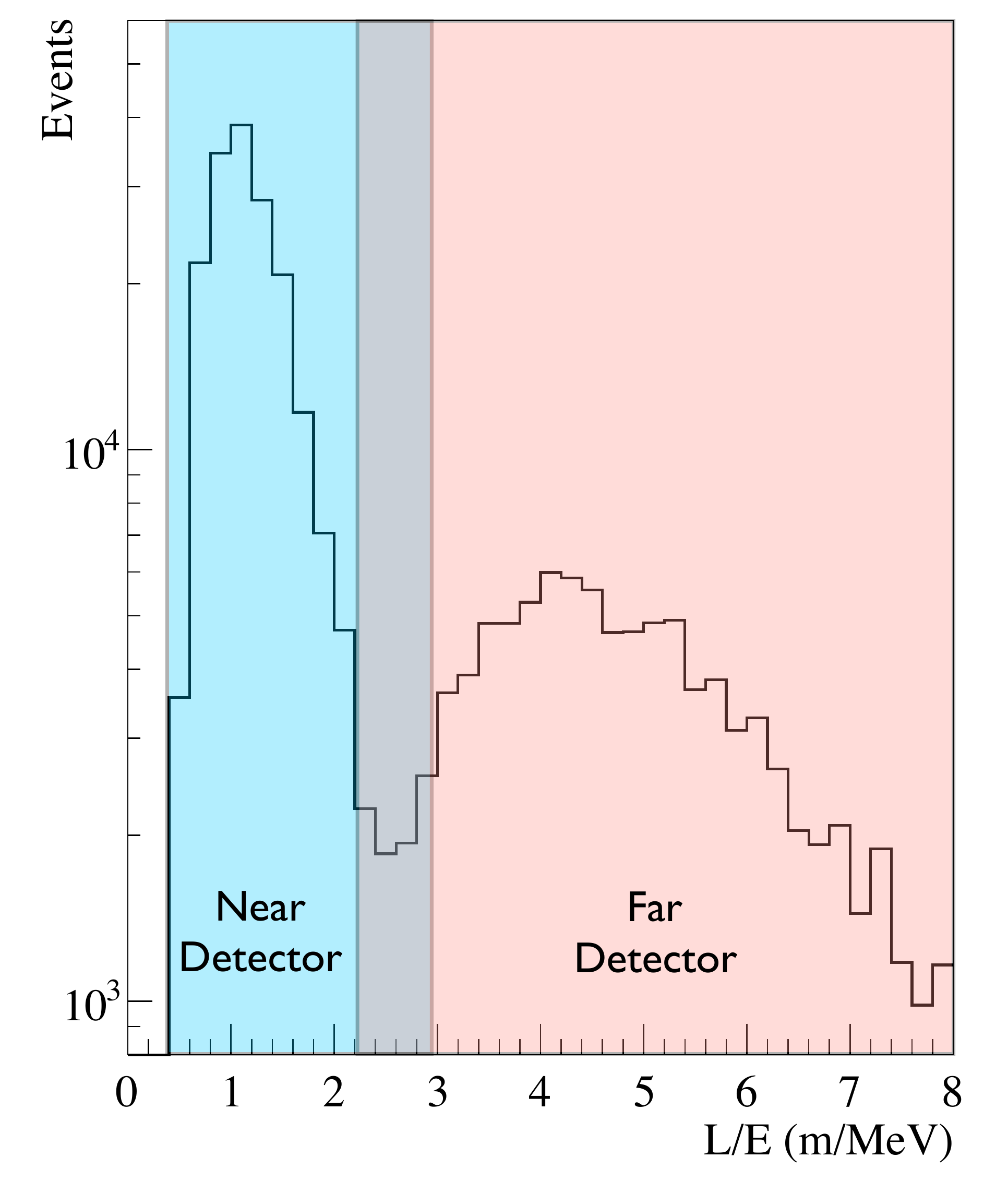}
\caption[]{Left: Sensitivity of the default experimental arrangement to 3+1 oscillations for different far detector distances. For the default two-detector arrangement, good sensitivity over a broad range of $\Delta m^2$ is achieved with $r_{min}$ around 11~m.  Above this, the range of sensitive mass splittings is further increased at the expense of a dip in sensitivity at mid-range mass splittings.  Right: $L/E$ distribution for the default two-detector arrangement and $r_{min}= 19$\,m for the far detector.  A dip is clearly visible at mid-range $L/E$ resulting in reduced sensitivity at mid values for $\Delta m^2$.
}
\label{fig:FarDist}
\end{figure}

The $L/E$ distribution for a multi-detector experiment depends on the minimum distance between the reactor and the near detector, the length of the detectors and the distance to the far detector's target region. We call the minimum distances to the near and far detectors $r_{min}^{near}$ and $r_{min}^{far}$ respectively.  Figure~\ref{fig:FarDist} shows the sensitivity of the default two-detector arrangement for various values of $r_{min}^{far}$.  For longer $r_{min}^{far}$, the sensitivity is highest for low and higher mass splittings of $\sim$0.3 eV$^2$ and above 2~eV$^2$, with a trough of low sensitivity in the middle-$\Delta m^2$ region around 0.7~eV$^2$.  As $r_{min}$ is decreased and the far detector is brought closer to the near detector, this trough disappears, along with a gradual shift of the low-$\Delta m^2$ sensitivity upwards, giving fairly consistent sensitivity over a broad range of mass splittings.  As $r_{min}^{far}$ is further lowered, sensitivity at low $\Delta m^2$ continues to degrade as sensitivities at mid- and high-range splittings begin to improve significantly.  

This behavior with varying far $r_{min}$ and detector length can be understood from the distribution of signal events in $L/E$ shown in Figure~\ref{fig:FarDist} for the default two-detector arrangement with near and far $r_{min}$ of 4~m and 17~m, respectively.  The peaks in this distribution corresponding to the near (low $L/E$) and far (high $L/E$) detector targets.  The relative lack of statistics in the mid $L/E$ range directly contributes to the lower sensitivity at medium $\Delta m^2$.  This also illustrates that if the detector lengths are altered, the widths of the peaks in $L/E$ change but never completely cover the dip in the sensitivity at medium $L/E$. As a result, the optimum near/far $r_{min}$ ratio will remain relatively consistent.  In general, near and far $r_{min}$ should be chosen to give continuous statistical coverage in $L/E$. The near-far distance ratio can be optimized to provide good sensitivity over a continuous, broad range of mass squared splittings taking into account the specific constraints and experimental parameters for the reactor sites under consideration.

\subsection{Far Detector Position Resolution}

\begin{figure}[htb!pb]
\centering
\includegraphics[trim=0.1cm 0.1cm 0.1cm 0.1cm, clip=true, width=0.55\textwidth]{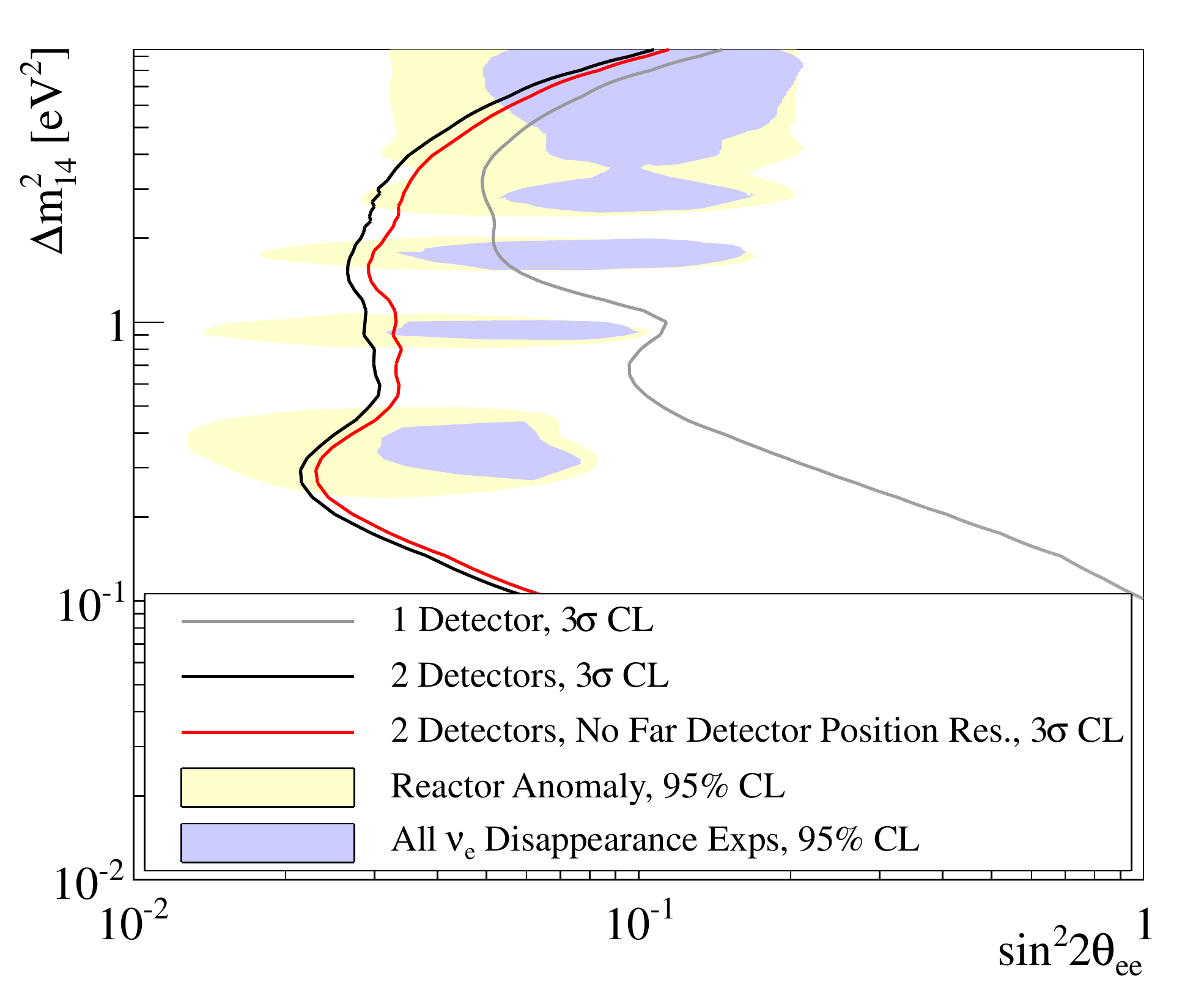}
\caption[]{Left: Sensitivity of the default experimental arrangement to 3+1 oscillations assuming differing position resolutions in the far detector.  A complete loss of far detector position resolution causes only a slight loss of sensitivity over a range of mass splittings.  A 2-detector experiment with no position resolution in the far detector still provides  a significant improvement in scientific reach over a single-detector experiment.}
\label{fig:FarRes}
\end{figure}

As discussed in Section~\ref{subsec:FarShape}, a far detector with small baseline range is best at optimizing the sensitivity of a multi-detector SBL oscillation experiment.  This conclusion suggests that, other than knowing the location of the far detector, event position information at the far site may not be needed.  We illustrate the value of far detector position information by increasing the resolution of the far detector up to the the total size of the active detector itself. The resulting sensitivity compared to the default case is shown in Figure~\ref{fig:FarRes}.  One can see that relatively little sensitivity is lost, still giving a large improvement over a 1-detector configuration.

Although less important for resolving L/E oscillations at the far site, good far detector position resolution may still be necessary at the far site for a number of reasons.  Sufficient far detector background reduction may require topology cuts, which would necessitate the use of position information.  Sufficient cancellation of relative systematics between all detector sub-segments may require that far and near detectors maintain the same segmented structure and position resolution.  Detailed simulations and background measurements are nessesary to quantify these effects.


\section{Conclusion}

The broadened experimental $L/E$ range afforded by an experiment with two multiple detectors significantly increases the sensitivity of a SBL experiment, particularly at lower values of mass-squared splitting. 
For the default experimental arrangement considered in this paper, 3$\sigma$ sensitivity to mixing amplitudes of 10\% or greater increases from 1-8 eV$^2$ for one detector to 0.1-10 eV$^2$ for two detectors.  This expanded range allows an SBL experiment to probe all the suggested mass-squared splitting ranges suggested by the the disappearance anomalies.  In addition to the improved  sensitivity in greater range in mass splitting, the overall sensitivity limit for a two-detector experiment is higher: the highest 3$\sigma$ sensitivity for 3+1 mixing is increased from 5\% for one detector to 2-3\% with an additional far detector of the type described in this paper.  This added sensitivity significantly improves coverage of the 95\%~CL suggested reactor anomaly parameter space.  
 Finally, the addition of a second detector improves a SBL reactor experiment's ability to distinguish between the existence of either one or two sterile neutrino states.

We have explored the experimental parameters for the second, far detector.  This detector should be larger than the near detector but it is unnecessary to achieve equal statistics between the near and far detectors.  The far detector length should be confined to a small range of baselines so as to provide the best normalization at larger distances, rather than increase the overall coverage of baselines.  Position resolution in the far detector is not critical: without any far position resolution, one can retain a majority of the sensitivity that one would have with identical near and far site position resolutions.  The distances between the near and far detectors should be optimized achieve a wide overall range in $L/E$ while avoiding gaps in $L/E$ coverage.  

\acknowledgements
We thank Henry Band, Randy Johnson, David Webber, and Nathaniel Bowden for helpful discussions.  This work was prepared with support from the Department of Energy, Office of High Energy Physics, under grants DE-FG02-95ER40896 and DE-FG02-84ER-40153, the University of Wisconsin, and the Alfred~P.~Sloan Foundation.

\bibliographystyle{revtex}
\bibliography{Reactorstudy}
\end{document}